\documentclass[12pt]{article}
\usepackage{amsmath}
\usepackage{amssymb}
\usepackage{epsfig}
\usepackage{cite}
\usepackage{color,colordvi}

\begin{document}
\vspace{0.01cm}
\vskip-0.5cm
\begin{center}
{\Large\bf  Strong Coupling and Classicalization}

\end{center}

\vspace{-0.1cm}

\begin{center}

{\bf Gia Dvali}$^{a,b,c}$\footnote{In part, based on lectures given at 
Erice summer school ``Future of Our Physics Including New Frontiers"  and at  LHC SKI 2016 conference.}

\vspace{.6truecm}


{\em $^a$Arnold Sommerfeld Center for Theoretical Physics\\
Department f\"ur Physik, Ludwig-Maximilians-Universit\"at M\"unchen\\
Theresienstr.~37, 80333 M\"unchen, Germany}


{\em $^b$Max-Planck-Institut f\"ur Physik\\
F\"ohringer Ring 6, 80805 M\"unchen, Germany}

{\em $^c$Center for Cosmology and Particle Physics\\
Department of Physics, New York University, \\ 
4 Washington Place, New York, NY 10003, USA}

\end{center}

\vspace{0.1cm}

\begin{abstract}
\noindent  
 {
\small
Classicalization is a phenomenon in which a theory prevents itself from entering into a strong-coupling regime, by redistributing the energy among many weakly-interacting soft quanta.  In this way, the scattering process of some initial hard quanta splits into a large number of soft elementary processes. In short, the theory trades the strong coupling for a high-multiplicity of quanta.  At very high energies, the outcome of such a scattering experiment is a production of soft states of high occupation number that are approximately classical.  It is evident that black hole creation in particle collision at super-Planckian energies is a result of classicalization, but there is no {\it a priory} reason why this phenomenon must be limited to gravity.  If the hierarchy problem is solved 
by classicalization, the LHC has a chance of detecting a tower of new resonances. The lowest-lying resonances must appear right at the strong coupling scale in form of short-lived elementary particles.  
The heavier members of the tower must behave more and more classically: 
they must be longer lived and decay into higher numbers of soft quanta.}

\end{abstract}

\thispagestyle{empty}
\clearpage

\section{Self-completion and classicalization} 

 The fundamental physics is about understanding nature at 
 different length-scales.   In effective field theory we formulate a description  in terms of some quantum degrees of freedom that are the most suitable ones for 
a given energy.  The degrees of freedom that we consider as suitable are 
{\it weakly-interacting}.  Each description has its domain of applicability,
 beyond which it breaks down and must be completed by a more powerful 
 description.  Hence, when moving towards shorter distances (high energies)
 we perform an UV-completion of the theory.  The signal for the need of UV-completion is that some of the degrees of freedom become strongly interacting 
above certain energy scale $\Lambda$.  
   In the standard approach, that we shall refer to as Wilsonian, the 
 UV-completion is achieved by means of integrating-in some new weakly-interacting 
 degrees of freedom. 
 
  The focus of this lecture will be an alternative - {\it non-Wilsonian } - approach,
  based on the ideas of self-completion \cite{selfcompletion} and 
  classcalization \cite{CLASS, CLASSGR, 2toN, CLASS3}. The key novelty is that the theory
 - instead of introducing new degrees of freedom  above the scale 
 $\Lambda$ -  selfcompletes by using the same ``old" low-energy degrees of freedom, but in the state of high-multiplicity.  Because of the high occupation numbers these states behave approximately-classically.  Hence, we can say that the theory self-UV-completes by classicalization.  That is, in classicalization the role of  UV degrees of freedom is played by 
 the collective excitations of IR  particles of high occupation number.   
 
  In the present lecture we shall consider this phenomenon and its role in the solution 
  to the Hierarchy Problem.  The hierarchy problem is the problem of 
  UV-sensitivity of the mass of a Brout-Englert-Higgs boson and we shall first explain its essence and the crucial role of gravity in it. Next, we shall introduce the idea of classicalization and its potential role in the solution of the hierarchy problem.  
   This solution relies on the fact that the  energy scale above which 
 an interaction of a given elementary particle (e.g., Higgs boson)  classicalizes,  represents the cutoff scale for the elementary particle mass. 
 
 One natural candidate for such a classicalizing interaction is gravity \cite{selfcompletion}.   However, there is no {\it a priory} reason 
why the role of the classicalizing interaction that stabilizes the Higgs mass must be limited to gravity.  
Not less interesting - and perhaps much more economical - possibility  would be to stabilize the Higgs mass via some classicalizing non-gravitational interaction of the Higgs particle, as suggested in \cite{CLASS}. The necessary condition for such a scenario is that the new interaction 
must become strong around TeV energies.   

  One of the messages of the present lecture is this: 
 any observational evidence of a new interaction of the standard model species, that becomes strong not far above TeV energies, 
should be considered as a potential signal of the solution to the hierarchy problem  
via classicalization.  In the latter case, above the strong-coupling scale we must observe a tower of massive resonances (``classicalons").  The  ``level of classicality" of these resonances  must be  an increasing function of their mass in the sense that the heavier resonances must be longer lived with their  
decay products being softer and of a higher multiplicity.

\section{Gravity makes the Hierarchy Problem real}

   There are  several motivations for new physics  beyond the Standard Model:  \\ 
   
  1) More simplicity, elegance and predictivity of possible extensions of the Standard Model; 
  
  2) The need to account for the phenomena of nature that the Standard Model cannot explain (e.g., dark matter, dark energy, inflation, baryogenesis);  

  3) The unification principle (e.g., unification of gauge forces and/or  unification with quantum gravity);  
  
  4) Naturalness. \\
  
   I shell split the naturalness problems into the following two categories: \\

  {\it I. Problems of UV-sensitivity.}  An example of this sort is the Hierarchy Problem:  the quadratic sensitivity of the Higgs (mass)$^2$ to UV-cutoff.  \\
  
  {\it II. Problems of Vacuum (Super)Selection.}   An example is the Strong-CP Problem originating from the $\theta$-vacuum in QCD. \\

   The hierarchy problem has a clear physical meaning because of gravity, or to be more precise, because of the existence of macro (classical)  as well as micro (quantum) black holes that it predicts.   
   
 \subsection{The two worlds: particles meet black holes}    
    
   Gravity contains a fundamental  quantum scale, the Planck length, defined in the following way,
      \begin{equation}
 L_P^2 \, \equiv \, \hbar G_N \, ,
 \label{LP}
 \end{equation} 
 where $G_N$ is the Newton's constant. The corresponding mass scale, 
 $M_P \equiv  {\hbar \over L_P}$, is the Planck mass. \footnote{We shall set the speed of light equal to one, but keep $\hbar$ explicit. We shall  ignore all the irrelevant numerical coefficients
 throughout the talk.}  
     
   The Planck mass represents a boundary between the world of elementary particles 
    and the world of black holes.  The elementary particles heavier than $M_P$ do not exist, they are black holes!
    
  In order to understand this, consider a particle of mass $m$.   There come the two associated length-scales with it.  The first one is the Compton wave-length, $L_C \equiv {\hbar \over m}$. 
 This length represents a distance at which the energy of quantum fluctuations exceeds the 
 mass of the particle.  The second length-scale is the gravitational radius, 
 $L_g \equiv \hbar {m \over M_P^2}$.    An object of mass $m$ localized beyond this length-scale becomes a black hole of Schwarzschild radius $L_g$ . Notice, this length-scale is {\it classical}, because it is  independent of $\hbar$.  
 We can now consider the following regimes.  \\
 
 { \bf The elementary particle regime:} \\
 
 For $m < M_P$,  we have, 
$L_C > L_P > L_g$, and thus, the Compton wavelength of a particle 
 is the dominant length-scale.  For example, if we think of a particle 
 of mass $m \ll M_P$ 
 as a gravitating source, we discover 
  that the classical Newtonian gravitational potential created by such a particle at distance $r \sim L_C$ is very weak, 
 \begin{equation}
  \phi_{Newton} \sim {mG_N \over L_C} = {L_g \over L_C} \, \ll \, 1 \, .
 \label{Nfield}
 \end{equation}
 This means that if we approach the particle from infinity, the quantum effects become important way before we have a chance to
 probe the scale $L_g$, i.e., the gravitational radius is completely shielded by quantum effects.  This is why the elementary particles - despite being treated as point-like - cannot be considered as black holes. \\
 
{\bf The black hole regime:} \\
 
 For  $m > M_P$, the story changes dramatically. We now have $L_C <  L_P <  L_g$ and the dominant length is the gravitational radius. 
  We are dealing with a black hole!  \\

 {\bf The meeting point: Planck mass black holes} \\
  
  The three length-scales meet,
$L_C = L_P = L_g$,  for  $m=M_P$. Hence, the Planck mass is an absolute  UV-cutoff in the sense that it represents an upper bound on the mass of any elementary particle.  
 
 The particles with $m\sim M_P$ are on the boundary of the two worlds and combine 
 properties of both. On one hand, they are strongly gravitating at their  Compton wavelength, i.e., the Newtonian potential (\ref{Nfield}) created by them at the distance $r \sim L_C$ is order one.  On the other hand, they are still far from being classical black holes, since quantum fluctuations are 100 percent important and cannot be ignored.  
 So,  I shall refer to such particles as {\it quantum black holes} (or 
 Planckions \cite{planckions}).
 
 It is very important to stress that  $M_P$ is not just an upper bound on the mass of the elementary particles, but particles of this mass actually exist in the spectrum of Einstein gravity,  since they can be explicitly ``manufactured" as the latest stage of 
 black hole evaporation and thus represent the inevitable part of the gravity spectrum 
 \cite{selfcompletion}. \\

 Before we continue with the main subject of the talk, let me make the following remark. 
  If a theory contains a significant number of particle species, $n$, the 
 domains of the above regimes are modified.  In particular, the mass of the lightest (quantum) black holes is 
 lowered from $M_P$ to ${1 \over \sqrt{n}}M_P$ \cite{species1}. Moreover, the quantum nature of 
 such black holes manifests itself also in the fact that their couplings to 
 different species cannot be universal \cite{species2}, i.e., 
 such a black hole will decay into different particle species (e.g., electrons and muons)  at different rates.  
 These are the important features that must be taken into the account in experimental searches of micro black holes. However,  for simplicity  of the present discussion we shall ignore this complication and assume the small number of species $n \sim 1$. \\

\subsection{The role of quantum black holes in the hierarchy problem}

 The above-discussed separation between the worlds of elementary particles and 
 black holes - and with quantum black holes occupying the boundary of the two worlds - makes the  hierarchy problem real. 
 
   The separation of the two worlds tells us that  Higgs cannot be heavier than $M_P$, but it by no means explains  
   why it is $17$ orders of magnitude lighter than $M_P$. 
   
%
 It is an easy question to answer, for example,  why Higgs is much lighter than any macroscopic object, e.g., a planet earth?   Although it may sound unusual, 
such a question would be fully legitimate in quantum field theory without gravity, since 
in such a theory effectively $M_P = \infty$ and elementary particles can 
be {\it arbitrarily heavy}.   Hence, in such a theory an elementary scalar of any  finite mass would demand an explanation. 

  However, with gravity with finite $M_P$ the answer is obvious: 
 the  Higgs particle  cannot be as heavy as the planet  earth, because 
 earth is mach  heavier than $M_P$ (earth mass is $\sim 10^{33} M_P$).   
  Such a heavy Higgs cannot be described as an elementary particle,  but instead would be a 
 classical black hole of size larger than a centimeter.  Thus, we know that non-perturbative gravity prevents the Higgs particle from
 correcting its physical mass - both  classically and  quantum mechanically - beyond $M_P$.  But, this leaves us with the question:  why is the Higgs mass-square about $34$ orders of magnitude smaller than the scale of its natural stability?

 Before continuing, we would like to close an imaginary loophole that questions the  reality of the hierarchy problem.  The argument goes as follows.  Since the hierarchy problem is about perturbative 
 sensitivity of the Higgs mass, it relies on the existence on a Wilsonian cutoff 
 scale to which the Higgs mass can be quadratically sensitive. 
 The role of such a cutoff can be played by some new heavy particles. 
 Therefore, one may argue that if the cutoff is absent, i.e., if there is no new Wilsonian physics beyond the standard model, the ground for the hierarchy problem becomes shaky.   
   Can this be the case?  The question hangs on whether such a cutoff may be  avoided in case if gravity is self-UV-complete in a non-Wilsonian way, 
as suggested it  \cite{selfcompletion}.   As also explained above, 
this self-completion happens via the phenomenon of  
 classicalization \cite{CLASS, CLASSGR, 2toN}, which is the central focus of the present lecture. This will be discussed in more details below.  
 Here we shall only mention that the key point of self-completion idea is that in deep UV gravity classicalizes, due to black holes.  In other words, at energies
 $ \gg M_P$  no new perturbative degrees of freedom are required for the consistency of the theory.   
 
 This may create a false impression that in such a case one is free to avoid an  UV-cutoff to which  the Higgs mass is quadratically sensitive.  This is {\it not}  the case: in the light of existence of quantum black holes of mass $M_P$, gravity provides an explicit source for quadratic sensitivity of the Higgs mass with respect to $M_P$. 
  
   As said above, we cannot exclude the quantum black holes of $M_P$-mass
  from the spectrum, since they can be explicitly produced as the latests stage of evaporation of macroscopic black holes \cite{selfcompletion} (see also \cite{transP}, and \cite{QBH3, QBH2} for recent discussions).  
  As pointed out in \cite{selfcompletion}, such quantum black holes contribute new massive poles at  $p^2 \sim M_P^2$ in the graviton propagator. Consequently, they  contribute into the renormalization of the Higgs mass at the loop level just as any other heavy massive particle would do.  
 The loop diagrams with such virtual quantum 
black holes of mass  $M_P$ are unsuppressed by any entropy factor and 
provide $\sim M_P^2$ corrections to the Higgs mass.   
 
  Putting everything together, we can summarize the Hierarchy Problem as a question: \\
  
 {\it  Why is the Higgs particle so far from being a quantum black hole (or, is it
 truly that far)? } \\

 As we shall discuss below, the answer to this question may lie in  a 
phenomenon of {\it classicalization}  \cite{CLASS, CLASSGR}.

 \section{Classicalization}

    If the hierarchy problem is not a problem of vacuum-selection, there must exist  some new physics around  TeV energies.  This new physics can be weakly-coupled or strongly-coupled.        
     Let me focus on a strongly-coupled case. 
 The usual attitude towards the strong coupling is that this is something we should be scared of.  I shall take a different attitude. First, 
the strong-coupling  is probably the best thing that could happen at LHC:  if LHC reaches the strong-coupling scale we cannot miss new physics.  But, this is not the only reason for my focus  on the strong coupling in this lecture.  The strong coupling will force us to think about new possibilities for UV-completion that until recently  have been overlooked. 
    
     Let me explain this.  In quantum field theory,  the strength of the interaction 
    among elementary degrees of freedom is controlled by  
 a quantum coupling, which can be chosen to be dimensionless and which 
 I shall denote by $\alpha$. 
 I shall assume that the coupling 
  is properly normalized in such a way that for $\alpha \ll 1$ the interactions are weak and the scattering rate is small.   For $\alpha \gtrsim 1$ the interaction becomes strong and 
  the naively-evaluated scattering amplitude violates unitarity.  
  In quantum  field  theory the value of the coupling $\alpha(E)$    
 depends on the energy scale $E$ at which it is evaluated.  Thus, the strength of the coupling 
 is a scale-dependent notion.   
 
   What happens when a theory hits the strong coupling,
  $\alpha(\Lambda) \sim 1$, 
    at some energy scale
   $\Lambda$? Above the scale $\Lambda$ the theory becomes a theory of something else: 
   new degrees of freedom enter the game. 
   Here we can distinguish the following three possible scenarios.  \\
  
 \subsection{UV-completion by addition of new species}  
   
   In the first scenario the new degrees of freedom can coexist with the old ones.
    That is, for $E < \Lambda$ the relevant  degrees of freedom are some $X$-particles, whereas 
    for $E > \Lambda$ the relevant degrees of freedom are  $X$-s and $Y$-s.  
     A well-known example is the Brout-Englert-Higgs  boson in the standard model. 
    
    In the standard model without Higgs, the interaction of the longitudinal 
    $W$-bosons becomes strong for energies above  the scale $\Lambda$  not far above the weak scale.
    This strong coupling is resolved by inclusion of the Higgs boson, which unitarizes the $W-W$ scattering 
    at high energies. All the low energy degrees of freedom of the standard model
    continue to exist as the legitimate weakly-interacting degrees of freedom also in the high-energy domain with the sole addition of the Higgs boson.  \\
    
  \subsection{UV-completion by the total  renewal of species}  
       
      In the second scenario the theory fully changes
   across the scale $\Lambda$: the old  IR degrees of freedom $X$ are
   no longer legitimate at energies $\gg \Lambda$
   and are replaced  
   by completely new UV degrees of freedom $Y$.   \\

    The example of the latter behavior is provided by QCD.  In  QCD the role of the strong coupling scale is assumed by the  QCD scale $\Lambda_{QCD}$.  
    For energies below $\Lambda_{QCD}$, or equivalently, for distances 
 much larger than $L_{QCD} \equiv \hbar \Lambda_{QCD}^{-1}$, the relevant degrees of freedom are
    mesons, glueballs and baryons.  (Despite the fact that some of these are heavier than 
    $\Lambda_{QCD}$, they are the correct asymptotic $S$-matrix states that an IR-observer can detect).   
    In contrast, at distances smaller than $L_{QCD} $ the good degrees of freedom are quarks and gluons.   \\

 In such scenarios the universal effect is that when we approach the scale $\Lambda$ from either side,  the degrees of freedom become strongly-coupled: 
 both pions as well as the quark and gluons become strongly-coupled at 
the QCD scale.   
  Due to this, the scale $\Lambda$ is an universal regulator in the sense that 
the weakly-coupled degree of freedom  ``inhabiting" a given domain cannot cross over into a domain of their strong coupling. 
It is this strong coupling that forces us to re-diagonalize Hamiltonian and 
change the basis from $X$-s to $Y$-s, whenever we cross the scale $\Lambda$. \\

\subsection{A self-UV-completion by soft multiplicity: classicalization}

   We shall now introduce the third scenario, in which UV-completion happens
   by {\it self-completion}: the UV theory employes the same IR degrees of freedom $X$, except in states with very large occupation  numbers. 
   
     For understanding the meaning of this phenomenon it is useful to think about gravity.     
 The case with gravity shares some similarities with QCD, but is fundamentally different in other aspects. 
In gravity, the role of $\Lambda$ is taken-up by the Planck mass $M_P$. 
  As we have discussed above, for energies below $M_P$ gravity represents a theory of graviton ( interacting with other elementary particle species), whereas way above the scale $M_P$, gravity becomes a theory of classical states, such as black holes. 
  In  particular, as we have discussed earlier, there exist no elementary particles 
  with the mass exceeding $M_P$.  
  
  This already indicates the dramatic difference between gravity 
  and the two above-discussed scenarios of UV-completion:  the macroscopic  black holes are not independent quantum particles, but rather 
multi-graviton states \cite{giacesar,critical, NP} \footnote{As noticed in \cite{giacesar}, in this limited sense, there exist some counterparts to black holes in QCD, in form of baryons.  For large number of colors, baryons are much heavier that the QCD scale, but by no means they represent  new UV-degrees of freedom. Rather, they are the composite multi-particle states.}.  In fact, strictly-speaking, in gravity 
all the states with the center of mass energy much above $M_P$ are the composite 
multi-particle states. 

  From the first glance, this may sound surprising, since one can easily  imagine a two-particle state at very large separation with arbitrary-high 
  center of mass energy $\sqrt{s} \gg M_P$.  Such a state can be obtained by separating the two particles (e.g., gravitons)  by a large distance 
  and  boosting them relative to each other. If the separation $r$ is much larger than 
 the gravitational radius corresponding to $\sqrt{s}$ (i.e., $r \gg \sqrt{s}G_N$)  such a state is not a black hole.  
  However,  all such states with center of mass energies exceeding $M_P$  either explicitly or 
 secretly consist of many quanta: gravitational field created by  
 these particles is equivalent of dressing them by many soft gravitons 
 \cite{CLASSGR, giacesar, 2toN, Ncount}.  If  separation $r$  is large, the dressing 
 gravitons are very soft and they are not contributing into the gravitational self-energy
 significantly.  However, their 
 presence is absolutely crucial for understanding the quantum nature of the state, especially 
 if the two initial particles are going to collide and form a black hole.   
  We shall come back to this 
 discussion below. \\

 For now, following \cite{CLASSGR, giacesar}, the crucial feature that we take from gravity is that the
 states of trans-Planckian energy are not dominated by new UV-degrees of freedom, but rather represents the states consisting of many soft IR-quanta.  
 This understanding of gravity, naturally leads us to the third scenario of UV-completion: 
 classicalization \cite{CLASS, CLASSGR}.  \\

The scenario of classicalization is a very peculiar case in which 
the role of $Y$-s is played by the collective excitations of many soft 
$X$-s.  \\

  Consider a theory with a four-point interaction of some IR degrees of freedom $X$, 
 \begin{equation} 
         \alpha X^4 \, ,
  \label{4X}
  \end{equation} 
 with the effective coupling $\alpha$ that gets strong 
 above the scale $\Lambda$.  Consider a head-on collision of two $X$-quanta with the 
 center of mass energy $\sqrt{s} \gg \Lambda$.   Then, the coupling evaluated at 
 such a high energy is strong,  $\alpha(\sqrt{s}) \gg 1$.  So,  the  unitarity seems to be violated. 
 This violation means that the perturbation theory in $\alpha$ breaks down. 
 If the theory is to make any sense, this perturbative expansion must be replaced by something else.  \\
 
 We shall now try to make a guess on how the theory could cure itself.  Our task is to use our imagination and design a scenario  
 that could avoid violation of unitarity, but the rules of the game are:  \\
 
  {\it 1) We are not allowed to invent new elementary quanta above the scale $\Lambda$;  \\
  
  2) We must respect all the basic rules of the quantum field theory (i.e., demand causality, conservation of energy-momentum, positivity of norm and energy, etc...)  
  } \\
  
    Following \cite{CLASS, CLASSGR}, our proposed solution is the following. 
    The task is to get the theory out of the strong coupling regime using its own IR resources.   
     The problem comes from  the fact that 
   the energy per $X$-quantum is too high and if this energy momentum will get exchanged among the two quanta 
   the quanta will be in a strong-coupling
  regime.  So, in order to avoid this, the system must turn the two-particle 
  scattering process
  $2X \rightarrow 2X$  into a multi-particle process 
  \begin{equation}
   2X \rightarrow NX \,, 
 \label{Nprocess}  
   \end{equation} 
 consisting of many elementary processes 
  in such a way that the momentum exchange per-quantum in each elementary process is small-enough
  so that the corresponding coupling  $\alpha$ is weak.  
   In this way, the system has a chance  to never leave the weak-coupling regime throughout the scattering process.  
  
     That is,  the entire energy $\sqrt{s}$ should be re-distributed among 
 $N$ quanta, such that  $\alpha(\sqrt{s}/N) < 1$.   
   Equivalently, the number  $N$ must satisfy 
   \begin{equation}
   \sqrt{s}/N < \Lambda \, , 
   \label{condN}
   \end{equation} 
  in order for the energy per quantum to be  
  below the strong coupling scale.  
     
 Obviously, higher is $\sqrt{s}$, more quanta must be produced in order to satisfy this requirement.    
  Hence,  such a theory  has a chance to  unitarize at the expense of creating 
  more and more soft quanta with an increasing center of mass energy.  But,  the states of high-occupation number 
  are behaving approximately {\it classically}. Hence,  the name {\it classicalization}
 \cite{CLASS,CLASSGR}.    
 
  Thus, classicalization is a phenomenon describing a situation in which  a theory prevents itself 
 from entering the strong-coupling regime at high energy, $\sqrt{s} \gg  \Lambda$ , by means of redistributing the high center of mass energy among $N$ soft quanta, such that  they satisfy (\ref{condN}) and therefore are weakly interacting.   
   
   As we shall explain later, this is precisely what happens in gravity in a very high energy particle collisions.   
    However, there is no {\it a priory}  reason why this phenomenon should be 
 restricted to gravity.   A theory with a genuine strong coupling scale  $\Lambda$ -
 and with no Wilsonian UV-completion available above this scale - can save unitarity by  classicalization.   
  
  \section{Hierarchy problem and classicalization}

  An interesting question - both from fundamental as well as the phenomenological points of view -  is whether the hierarchy problem can be solved 
  via classicalization due to some new interaction of the  standard model 
  Higgs?
  This could happen,  if the Higgs particle is subjected to some fundamental interaction    
   that becomes strong above a cutoff scale $\Lambda \sim$ TeV.  
    Let us focus on a situation 
  when this new classicalizing  interaction does not imply introduction of new elementary particles \cite{CLASS,CLASSGR}. 
   
  In such a case we have to endow the Standard Model sector with interaction that becomes strong around TeV energies, but we must forbid the resolution of this 
  interaction by some new weakly-coupled degrees of freedom.  In other words, the only possibility for the system to unitarize is due to production of multi-particle states composed out of the low energy quanta of the Standard Model.    
  
   For example,  let us consider modeling the strong coupling effect in UV
  by some simple derivative self-interaction of the Higgs boson, 
    \begin{equation} 
    G \,  (\partial_{\mu} H \partial^{\mu}H)^2 \,   =\,  {1 \over \hbar} { 1 \over \Lambda^4} \, (\partial_{\mu} H \partial^{\mu}H)^2 \, .
 \label{self}
    \end{equation} 
   (The gauge structure, which plays no role in this discussion, has been ignored).  
   Here $G$ is the coupling constant of a {\it classical} theory, 
   which has a dimensionality $[G] = {(length)^3 \over energy}$.  We have
   rewritten it in terms of a quantum cutoff scale as $G = {\hbar^3 \over \Lambda^4}$. 
   In  $\hbar \rightarrow 0$ limit, $\Lambda \rightarrow 0$ in such a way that 
   $G$ is finite, as it should.  The dimensionless {\it quantum}  coupling 
  for a characteristic  momentum exchange $p$,    
   is defined as 
   \begin{equation}
   \alpha \equiv   \hbar^{-3} G p^4 =  {p^4 \over \Lambda^4} \, . 
   \label{quantum}
   \end{equation}  
 
   We shall assume that 
   the above interaction is {\it fundamental}, in the sense that it is not obtained via integrating-out some weakly-coupled elementary degrees of freedom above the scale 
   $\Lambda$. 
   To put it differently, we demand that the strong coupling is {\it genuine}. 
   
   The above interaction 
   gives $2 \rightarrow 2$ scattering of the Higgs  bosons, with the naive effective coupling strength that grows as $\alpha \sim {s^2 \over \Lambda^4}$.  If system classicalizes, 
 then at energies $\sqrt{s} \gg \Lambda$, we expect the scattering to be dominated 
 by production of many soft Higgs quanta. Can we estimate their number $N$? 
  
  It is interesting that this number can be deduced in a model independent way, 
 from the requirement that the characteristic wavelength of the soft quanta is independent of  $\hbar$.
 As it was noticed  in \cite{CLASS}, there exists an unique classical length-scale that can be built out of the center of mass energy $\sqrt{s}$ and the coupling constants of the theory and it is given by, 
 \begin{equation}
   L_{cl} \,  \equiv \, {\hbar \over \Lambda} \left ( \sqrt{s} \over \Lambda \right )^{1 \over 
   3}  \, = \, (G \sqrt{s})^{1\over 3} \, .
 \label{radius} 
 \end{equation}
 In \cite{CLASS} this scale was called a {\it classicalization radius} (notation used there was $r_*\equiv L_{cl}$). 
   From here we can easily derive the number of quanta of the above wavelength 
  among which the initial energy $\sqrt{s}$ must be redistributed. 
   We get \footnote{Counting has to be modified once $L_{cl}$ exceeds
   the Compton wave-length of the Higgs particle, since the maximal number of 
   quanta produced is bounded by $N_{max} = {\sqrt{s}\over m_H}$.}, 
   \begin{equation}
     N \, = \, \sqrt{s} {L_{cl} \over \hbar} \,  = \left ( {\sqrt{s} \over \Lambda} \right )^{{4\over 3}} \, .
     \label{Nsoft}
     \end{equation}
    In the same time evaluating the effective coupling $\alpha$  given by (\ref{quantum})  for the soft quanta of wavelength 
    (\ref{radius}), we discover, 
     \begin{equation}
     \alpha \, = \,  \left ( {\sqrt{s} \over \Lambda} \right )^{-{4\over 3}}  \, , 
     \label{Nsoft}
     \end{equation}
or equivalently, 
\begin{equation}
     \alpha N = 1 \, .
   \label{criticalpoint}
   \end{equation}
    To put it shortly, we discover that the requirement that the wavelength  
  of the soft quanta be given by the unique classical scale existing in the problem, fixes the number of quanta 
  to be given by the inverse of their quantum coupling! 
  
  What is the significance of this fact? The answer is:  the quantum criticality of  
the system  of $N$ soft bosons \cite{critical,gold}. 
   This quantum critical point marks the state of the system for which 
  the collective interaction among $N$ soft bosons starts to be important: 
  despite the fact that the coupling among the individual quanta $\alpha$ is weak, the collective coupling, measured by $\alpha N$, is equal to one.     
 The studies of the simplest prototype systems of $N$-bosons  \cite{critical,gold}  
 show that at quantum criticality non-perturbative collective effects become extremely important.   In other to reliably determine viability of classicalization mechanism for the UV-completion, a non-perturbative input from
many-body physics is probably crucial.  In particular, this input  allows to 
eliminate the unitarity violating kinematical domain of the scattering process
\cite{2toN}.

    \section{Lesson from gravity}

     One existing theory for which we have a strong evidence indicating that  classicalization works, is gravity
   \cite{selfcompletion, CLASSGR, 2toN}. This is because the creation of multi-particle states 
    is guaranteed by the fact that the high-energy scattering 
 in gravity is dominated by black holes.  It is a well-known idea that in gravity in very high energy particle collisions black holes are  produced 
 \cite{BH1, BH2, BH3}.  After these pioneering papers, a lot of evidence has been gathered indicating that black holes are indeed formed in trans-Planckian  
scattering. \\ 
 
 However, the idea of classicalization sheds a new light at this phenomenon: 
 it offers a microscopic description of the process of black hole formation in particle collision in terms of the high-multiplicity graviton amplitude \cite{2toN}. \\

  In order to explain this, let us first recall the standard qualitative argument for black  hole formation in  high energy scattering, which goes as follows.  Consider a  collision of two elementary particles  at trans-Planckian  center of mass energy,  $\sqrt{s} \,  \gg \, M_P$.   We shall assume that there is no long-range gravity-competing repulsive force acting between these particles.  Then, if the impact parameter is less than the gravitational radius corresponding 
  to the center of mass energy, $L_g = \sqrt{s}L_P^2$,  the initial energy will get localized within the region smaller than its own gravitational radius.  Hence a black hole of  mass $\sqrt{s}$  is expected to form.    
  
  Following \cite{CLASSGR, 2toN},   let us now explain that - from the microscopic point of view - the  black hole creation in high energy particle collision is
classicallization at work.   According to this description, 
 from the point of view of a microscopic theory, 
   the amplitude capturing the essence of black hole formation is 
   a $2\rightarrow N$ transition of the two high center of mass energy  quanta into $N$ soft gravitons, of momenta 
  $\hbar /L_g$.   Correspondingly, their  number is given by $N = s L_P^2$.  
 Note that for this process we have $N = \alpha^{-1}$, where  
  $\alpha\, = \, M_P^2 /s$, is the gravitational coupling of final gravitons.  
  This is exactly the same relation as described by (\ref{criticalpoint}).  Thus, 
  from quantum physics point of view, the black holes are formed in the kinematic regime in which the produced $N$ soft gravitons obey the quantum criticality relation
(\ref{criticalpoint}).

   The rate of the $2\rightarrow N$ process in this kinematic regime 
  has been computed in \cite{2toN} and it scales as,  
  \begin{equation}
   \Gamma_{2\rightarrow N} \, \sim \,   \alpha^N N!  \sim e^{-N} \,,
 \label{regime} 
 \end{equation}
  where we have used Stirling's formula in order to capture the leading-order 
  exponential scaling, ignoring the power-law pre-factors and numerical coefficients.  
  
  Few observations are in order. First, notice that 
  the number of gravitons  $N$ of the wavelength $L_g$ - that can account for  the 
  initial center of mass energy $\sqrt{s}$ - scales as the Bekenstein entropy \cite{Bek}  
  of the corresponding mass black hole, 
 \begin{equation}
    N = \alpha^{-1} = S_{Bek}   \, .
 \label{entropy} 
 \end{equation} 
    Consequently, the exponential  suppression factor in (\ref{regime}) 
    matches the expected entropy suppression factor $\sim e^{-S_{Bek}}$ that one could 
    get - from very general semi-classical considerations - for  a  transition rate from an initial two-particle state 
    to any given black hole micro-state.  
  
   Secondly, notice that the black hole entropy scales as the inverse coupling 
   of the final soft gravitons, $S_{Bek} \sim \alpha^{-1}$.  
   
    From here we are learning the two things. 
    
     First, we clearly see glimpses  of 
  black hole entropy in  $2\rightarrow N$ amplitude.  Secondly, we see that this glimpses 
  appear in the kinematic regime in which the number and the coupling of outgoing gravitons satisfy
  the relation (\ref{criticalpoint}). 
%
    According to the black hole  portrait  of \cite{giacesar,critical, NP} this is not an accident. 
     According to this theory, the microscopic significance of the  relation (\ref{criticalpoint})  is that it corresponds to 
    the critical point of quantum phase transition, at which  $N$ gravitons 
    form a self-sustained bound-state, a black hole.  
 (Glimpses of this $N$-graviton constituency can be captured also by other considerations, such as, e.g., the count given in \cite{Ncount})).     
     Because of quantum criticality, 
    this point is characterized by appearance of collective  Bogoliubov  
    modes, with the tiny energy gap $\sim 1/N$. These are the modes that carry black hole entropy.  They result into an exponentially-large number of nearly-degenerate micro-states. 
   It is this quantum criticality that defies the usual intuition that creation of multi-particle classical configurations in two-particle collision  are expected to be exponentially-suppressed. 
   
    Before continuing,  let us note that the emergence of the same  critical number $N$ was recently observed in a seemingly-unrelated computation of soft graviton emission in trans-Planckian scattering \cite{Gabriele}, as well as in  eikonal approximation of $2\rightarrow 2$ graviton scattering \cite{FlorianBo}. 
    These results strengthen the evidence for multi-graviton nature of the black hole formation process.   
   
     Can we generalize the lesson from gravity to other strongly-coupled theories? 
    If yes, the guideline would be to look for quantum-criticality of $N$-particle state. 
     We saw that the regime of classicalization in non-gravitational scalar theory satisfies exactly the same relation (\ref{criticalpoint}) as the gravitons do in black hole formation process.  This may indicate an intrinsic connection between classicalization and quantum criticality.  
    
     One rather generic case when such a criticality is achieved is for the systems of  attractive bosons.  In such a situation, similarly to black hole formation case, 
     the critical point marks the regime in which the collective attraction of 
     $N$-bosons is strong enough for forming a long-lived bound-state.      
  The quantum critical point  $N\alpha = 1$ produces 
    a near-gapless excitations and correspondingly a  high-density of states. 
    Moreover, if the interaction among bosons is momentum-dependent the 
  number of  modes can scale as the power-law with $N$ and correspondingly 
  the number of states can scale exponentially with $N$ \cite{gold}. 
    This may produce an exponentially-large density  
     of micro-states, just as in black hole case, that could potentially compensate 
     the naive suppression factor of production of multi-particle states.   
   
 \section{Experimental prospects}   
     
       We finish the discussion by briefly  commenting on generic experimental signatures of classicalization. 
      We are learning that a classicalizing theory in a high-energy scattering process above the scale $\Lambda$ is dominated 
   by the states with many soft quanta.  These states represent the collections 
  of quanta of the softness ${\hbar \over  L_{cl}}$ and the occupation number $N$ given by 
  the relation (\ref{criticalpoint}) in which the coupling $\alpha$ has to be evaluated 
 for the momentum exchange ${\hbar \over L_{cl}}$.  Correspondingly, the number 
 of soft constituents $N$ grows with the mass of the state. 
   The precise scaling of $N$ with energy is set by the momentum-dependence of the coupling 
  $\alpha$, but the relation (\ref{criticalpoint}) is expected to be universal. 
  
  In theories in which the classicalizing strong interaction is attractive, 
  these multi-particle states are expected to be describable as soft bound-states.  
 They are soft in the sense that the characteristic wave-lengths of the constituents are of the same order as the classical size of the bound-state $L_{cl}$.  The bound-states are   
 expected to be unstable and after some time decay into $N$-quanta, via the process reminiscent of the  Hawking evaporation of black holes.  The decay-time is expected to increase with increasing $N$, so that the heavier bound-states are softer and are longer lived.  
  In the light of the fact that black hole formation in high-energy scattering in gravity is a particular example of classicalization, this analogy is not at all surprising.    
  
   Thus, if the theory classicalizes above some strong coupling scale $\Lambda$,
  experimentally we should observe production of the tower of resonances that starts above the center of mass energy $\sqrt{s} = \Lambda$. 
 If the hierarchy problem is solved by classicalization,  the strong coupling  scale $\Lambda$ should not be very far from  TeV energies.  
  
   The resonances 
   produced right at the threshold energy $\Lambda$ are special in the 
 same sense as the Planck mass quantum black holes are in gravity:
 they represent the boundary between the worlds of elementary particles and classicalons.      
 For these resonances the number of constituents $N \sim 1$. Correspondingly, both, the four-point coupling $\alpha$ as well as the collective coupling $\alpha N$ are order one.   This is very different from the bound-states produced at $\sqrt{s} \gg \Lambda$, since for them 
   $N \gg 1$ and correspondingly the coupling $\alpha \ll 1$.  
  Thus, the lightest resonances - of mass $\sim \Lambda$ - carry the hybrid features. On one hand they can be viewed as the bound-states 
 of two (or few) quanta of momentum $\sim \Lambda$. On the other hand - since 
 such quanta are in strong coupling regime - the bound-state can equally well be 
 described as a new fundamental degree of freedom, similarly to Planck mass quantum black holes in gravity \cite{selfcompletion}.  For the heaver 
   states the composite nature is much more apparent,  since the constituent quanta 
   are soft and coupling  $\alpha$ is weak.

   Thus, a generic experimental signature of classicalizing theories is
  a production of a tower of resonances (classicalons) above the scale $\Lambda$
  \cite{CLASS, CLASSGR, CLASS7}.  
  The lightest 
   resonances are expected to be short-leaved and to decay into very few 
 energetic quanta.  At any given center of mass energy, 
        $\sqrt{s} \gg \Lambda$,  the scattering is expected to be dominated by the production of the heaviest energetically-allowed  resonances. The heavier resonances  should be longer lived and decay into larger 
number of the soft quanta. \\
       
       It would be good to understand, whether there is any  connection between classicalization in a strongly coupled theory and the break-down of  perturbative computations in ordinary weakly-coupled theory in multi particle amplitudes 
\cite{multi}.  By some authors this growth of perturbative amplitudes has been interpreted as the source for unsuppressed production of many particles around $100$ TeV energies (see, \cite{multi2} and references therein). 

   From the first glance, the two phenomena are totally disconnected and, in some sense, are exact opposites. Indeed, in  classicalizing theories it is the four-point interaction that  becomes strong and a naively-evaluated $2\rightarrow 2$ scattering violates unitarity. 
The multi-particle $2\rightarrow N$  amplitude is precisely  what restores unitarity
for large-enough $N$.
 Hence, in this case the theory tells us that strongly-interacting hard particles are  
no longer the legitimate degrees of freedom and must be replaced by weakly-interacting 
{\it collective}  degrees of freedom of soft $N$-particle states.  

  In \cite{multi, multi2}, the story is exactly the opposite: the four-point coupling is perfectly weak and instead the $2\rightarrow N$ amplitude violates unitarity for large enough $N$.  So, in this case, the soft  multi-particle states are the ones with 
a questionable legitimacy.   

  Nevertheless,  due to a multi-particle nature of both phenomena,  the potential 
 relation needs to be better understood. \\
         
  We finish this note with the remark about the prospects of studying classicalizing theories at the accelerators of next generation. Even if LHC  reaches the bottom of the tower of resonances predicted by the classicalization solution of the Hierarchy Problem,  it most likely will not be able to probe the higher-mass resonances that are longer-lived and decay into higher number of soft quanta.  Probing these higher mass 
  resonances is necessary for concluding with certainty that we are indeed dealing with the classicalization phenomenon.  In this respect classicalizing theories have an advantage:
 the property of growing  cross-section 
 at high energies makes them into viable candidates for being probed 
 at the high energy accelerators even with relatively low luminosity, 
 such as discussed in \cite{Allen}.

\section*{Acknowledgements}
 
 I thank Cesar Gomez for valuable discussions and comment. 
 It is a pleasure to thank Prof. Antonino Zichichi for exciting discussions and for  invitation to Erice summer school "Future of Our Physics Including New Frontiers", as well as to thanks the organizers of LHC SKI 2016 conference for invitation and a stimulating feedback.

\end{document}